\documentclass[aps,twocolumn,prl,floatfix,showpacs,citeautoscript,superscriptaddress]{revtex4-1}

\usepackage[none]{hyphenat}
\usepackage{amssymb}
\usepackage{amsmath}
\usepackage{graphicx}
\usepackage{bm}
\usepackage{times}
\usepackage{color,soul}
\usepackage[dvipsnames,svgnames,table]{xcolor}
\usepackage{hyperref}
\usepackage{fancyhdr}
\usepackage{enumitem}
\usepackage[english]{babel}
\usepackage[caption=false]{subfig}
\usepackage{braket}
\usepackage{scalerel}
\usepackage{graphicx}
\usepackage{mathrsfs}
\usepackage{MnSymbol}
\usepackage{lipsum}

\hypersetup{
pdfstartview={FitH},% fits the width of the page to the window
colorlinks=true,    % false: boxed links; true: colored links
linkcolor=NavyBlue, % color of internal links
citecolor=Blue,   % color of links to bibliography
filecolor=NavyBlue, % color of file links
urlcolor=NavyBlue   % color of external links
}

\graphicspath{{/Users/nicolodefenu/Dropbox/generalized_higgs/Figs}}

\begin{document}

\title{
Engineering Higgs dynamics by spectral singularities
%  Higgs engineering in Bardeen-Cooper-Schrieffer systems
%Density-of-states engineering of Higgs mode dynamics in BCS systems
}

\author{H. P. {Ojeda Collado}}
\email{hector.pablo.ojedacollado@roma1.infn.it}
\affiliation{ISC-CNR and Department of Physics, Sapienza University of Rome, Piazzale Aldo Moro 2, I-00185, Rome, Italy}
\author{Nicol\`o Defenu}
\affiliation{Institut f\"{u}r Theoretische Physik, Eidgen\"{o}ssische Technische Hochschule Z\"{u}rich, 8093 Z\"{u}rich, Switzerland}
\author{Jos\'{e} Lorenzana}
\email{jose.lorenzana@cnr.it}
\affiliation{ISC-CNR and Department of Physics, Sapienza University of Rome, Piazzale Aldo Moro 2, I-00185, Rome, Italy}

\begin{abstract}
  We generalize the dynamical phase diagram of a Bardeen-Cooper-Schrieffer condensate, considering  attractive to repulsive, {\it i.e.},  critical quenches (CQ) and a non-constant density of states (DOS). We show that different synchronized Higgs dynamical phases can be stabilized, associated with singularities in the density of states (DOS) and different quench protocols. In particular, the CQ can stabilize an overlooked high-frequency Higgs dynamical phase related to the upper edge of the fermionic band.  For a compensated Dirac system we find a Dirac-Higgs mode associated with the cusp singularity at the Fermi level, and we show that synchronized phases become more pervasive across the phase diagram. The relevance of these remarkable phenomena and their realization in ensembles of fermionic cold atoms confined in optical lattices is also discussed.
\end{abstract}

\date{\today}
\maketitle

%%%%%%%%%%%%%%%%%%%%%%%%%%%%%%%%%%%%%%%%%%%%%%%%%%%%%%%%%%%%%

%%%%%%%%%%%%%%%%%%%%%%%%%%%%%%%%%%%%%%%%%%%%%%%%%%%%%%%%%%%%%

%Several platforms to simulate BCS optics, cavities..., ultra cold fermions...
%In previous work....

{\em Introduction.} Many-body systems are characterized by the occurrence of correlated phenomena, which have no counterpart in the few-body realm. In this perspective, the spontaneous symmetry breaking (SSB) of any Hamiltonian symmetry by the establishment of a finite order parameter represents one of the fundamental examples\,\cite{Nambu1961I,Sachdev2011,DiCastro2015,Nishimori2011}. Consequences of SSB include the appearance of superfluid and superconducting phases in condensed matter systems, as well as the occurrence of a finite mass of the intermediate vector bosons, the carrier particles of the weak force in the Standard Model\,\cite{Anderson1963, englert1964broken, higgs1964broken, guralnik1964global}. 

The excitations on top of the SSB ground state, in systems with continuous (gauge) symmetries, consists of Nambu–Goldstone (phase) modes and massive Higgs (amplitude) modes. In the Standard Model, the latter manifest themselves as the Higgs boson, whose experimental observation led to the 2013 Nobel Prize in physics\,\cite{kibble2009englert,alvarez2011eyes}.

The historical tight relationship between condensed matter and high-energy physics is rooted in the universality of continuous SSB transitions, whose appearance in fermionic systems has been first described by the paradigmatic Bardeen-Cooper-Schrieffer (BCS) theory\,\cite{cooper1956bound,bardeen1957theory}. Thus, it shall not surprise that Higgs mode dynamics has been observed across multiple systems in condensed matter, including superconductors with charge order\cite{Sooryakumar1980,Balseiro1980,Littlewood1981,Cea2014,Grasset2018}, quantum antiferromagnets\,\cite{ruegg2008quantum}, He$^{3}$ superfluids\,\cite{Halperin1990}. Its search in 
superconducting systems\,\cite{Matsunaga2013,Matsunaga2014,sherman2015higgs,Katsumi2018,Chu2020,Shimano2020} has led to intensive theoretical investigations\,\cite{Balseiro1980,Littlewood1981,Cea2014,Cea2015,Cea2016c,gazit2013fate,Silaev2019,Murotani2019}. Given the relevance and ubiquity of the Higgs mechanism, its observation was also the focus of  quantum simulations in the superfluid/Mott-insulator transition of lattice bosons\,\cite{bissbort2011detecting,Endres2012}, in spinor Bose-Einstein condensates (BEC)\,\cite{hoang2016adibatic} and in cavity-QED experiments\,\cite{Leonard2017}.

While the Lorentz invariance stabilizes the Higgs mode in the high-energy context, its signatures at low-energies are less sharp and crucially depend on the strength of the interactions\,\cite{podolsky2011visibility,scott2012rapid,Barlas2013,Cea2015,liu2016evolution,han2016observability}. A comprehensive picture of the Higgs mode features across the BEC-BCS crossover has recently been  obtained in a fermionic cold atom ensemble\,\cite{behrle2018higgs}, reinvigorating the interest of the cold atom's community in the signatures of SSB and Higgs mechanism also in the few-body limit\,\cite{bjerlin2016few,bayha2020observing}.

Upon temporal variation of a control parameter, many-body dynamics may display several peculiar features, which are reminiscent of the behavior of thermodynamic functions at transition points\,\cite{Heyl2018}. In particular, theoretical investigations uncovered the appearance of dynamical phase transitions following interaction quenches in strongly correlated systems\,\cite{Eckstein2009,garrahan2010thermodynamics,diehl2010dynamical,Schiro2010,sciolla2010quantum}. These dynamical transitions occur both as order parameter modulations and as singularities in the L\"oschmidt echo dynamics\,\cite{heyl2013dynamical,heyl2015scaling}, which have been observed in quantum optics experiments\,\cite{jurevic2017direct, zhang2017observation}. These two phenomena have been shown to be deeply intertwined both between each other and with the existing equilibrium transitions\,\cite{zunkovic2018dynamical,halimeh2017dynamical}, except for few examples\,\cite{halimeh2017prethermalization,defenu2019dynamical,uhrich2020out}.

Following the current perspective, we consider a critical quench (CQ) of the BCS model, {\em  i.e.}, an attractive to repulsive interaction quench where the sign of the coupling constant is flipped. Then, the system, which is prepared in the superconducting equilibrium state for attractive interaction, evolves according to a repulsive Hamiltonian, whose equilibrium ground-state would be a normal (non-superconducting) gas. Thanks to the flexibility of our approach, we can target a wide range of different models parametrized by diverse density of states (DOS).

The CQ protocol displays the distinctive features of the Higgs mode dynamics found in pioneering works~\cite{Barankov2004,Barankov2006a},  including synchronized oscillations of the order parameter.
However, contrary to the common belief, there is no a unique synchronized Higgs phase (SHP) but different SHPs
can be stabilized depending on the model and the protocol.
The oscillation frequency can be determined by  any spectral singularity.  This includes those singularities not connected with the SSB but present in the bare DOS. Thus, a combination of protocol and the optical lattice in a cold atom system allows engineering generalized  synchronized Higgs phases on-demand.

%%%%%%%%%%%%%%%%%%%%%%%%%%%%%%%%%%%%%%%%%%%%%%%%%%%%%%%%%%%%%
%{\em Out-of-equilibrium BCS model.}
{\em Model.} In order to prove our picture, we consider a weak-coupling fermionic condensate with $s$-wave pairing described by the BCS model and subject to a sudden quench of the pairing interaction. The Hamiltonian can be written as
%%%%%%%%%%%%%%%%%%%%%%%%%%%%
\begin{equation}
\label{eq:HBCS}
H_{\mathrm{BCS}}=\sum_{\bm{k},\sigma}\xi_{\bm{k}}\hat{c}_{\bm{k}\sigma}^{\dagger}\hat{c}_{\bm{k}\sigma}^{}-\lambda(t)\sum_{\bm{k},\bm{k^{\prime}}}\hat{c}_{\bm{k}\uparrow}^{\dagger}\hat{c}_{\bm{-k}\downarrow}^{\dagger}\hat{c}_{\bm{-k^{\prime}}\downarrow}^{}\hat{c}_{\bm{k^{\prime}}\uparrow}^{},
\end{equation}
%%%%%%%%%%%%%%%%%%%%%%%%%%%%
where $\xi_{\bm{k}}=\varepsilon_{\bm{k}}-\mu$ measures the energy from the Fermi level $\mu$ and the pairing interaction 
$\lambda(t)=\lambda_i\theta(-t)+\lambda_f\theta(t)$ with $\theta$ the Heaviside step function. Here $\hat{c}^\dagger_{\bm{ k}\sigma}$ ($\hat{c}_{\bm{k}\sigma}$) is the usual creation (annihilation) operator for fermions with momentum $\bm{k}$ and spin $\sigma$.

Due to the all-to-all interaction assumed in the last term of Eq.~(\ref{eq:HBCS}), a mean-field approach becomes exact in the thermodynamic limit. Thus, we shall consider the BCS mean-field Hamiltonian which can be written, using the Anderson pseudospin formulation~\cite{Anderson1958} as
%%%%%%%%%%%%%%%%%
\begin{equation}\label{eq:hamMF}
\hat{H}_{\mathrm{MF}}=-\sum_{\bm{k}}\hat{\bm{S}}_{\bm{k}}\cdot\bm{b}_{\bm{k}}.
\end{equation}
%%%%%%%%%%%%%%%%%
Here, $\bm{b}_{\bm{k}}\left(t\right)=(2\Delta\left(t\right),0,2\xi_{\bm{k}})$ represents an effective magnetic field vector for the $\frac{1}{2}$-pseudospin operator $\hat{\bm{S}}_{\bm{k}}=(\hat{S}_{\bm{k}}^{x},\hat{S}_{\bm{k}}^{y},\hat{S}_{\bm{k}}^{z})$ where
%%%%%%%%%%%%%%%%%%%
%\begin{eqnarray}
%\nonumber
$\hat{S}_{\bm{k}}^{x}=\frac{1}{2}\left(\hat{c}_{\bm{k}\uparrow}^{\dagger}\hat{c}_{-\bm{k}\downarrow}^{\dagger}+\hat{c}_{-\bm{k}\downarrow}^{}\hat{c}_{\bm{k}\uparrow}^{}\right)$\,,
$\hat{S}_{\bm{k}}^{y}=\frac{1}{2i}\left(\hat{c}_{\bm{k}\uparrow}^{\dagger}\hat{c}_{-\bm{k}\downarrow}^{\dagger}-\hat{c}_{-\bm{k}\downarrow}^{}\hat{c}_{\bm{k}\uparrow}^{}\right)$\,, %\label{eq:sz}
$ 
\hat{S}_{\bm{k}}^{z}=\frac{1}{2}\left(1-\hat{c}_{\bm{k}\uparrow}^{\dagger}\hat{c}_{\bm{k}\uparrow}^{}-\hat{c}_{-\bm{k}\downarrow}^{\dagger}\hat{c}_{-\bm{k}\downarrow}^{}\right)$. %\nonumber
%\,,
%\end{eqnarray}
%%%%%%%%%%%%%%%%% 
Without loss of generality, we have assumed that the equilibrium BCS order parameter $\Delta$ is real, which remain true over time due to electron-hole symmetry. The instantaneous BCS order parameter is given by 
%%%%%%%%%%%%%%%%%%%%%%%
\begin{equation}
\label{eq:deltat}
\Delta(t)=\lambda(t)\sum_{\bm{k}} S_{\bm{k}}^{x}\,,
\end{equation}
%%%%%%%%%%%%%%%%%%%%%%%%%
where symbols 
%$S_{\bm{k}}^{x}$,
without hat denote the expectation value of operators
%$\hat{S}_{\bm{k}}^{x}$
in the time-dependent BCS state. % (this notation is used hereafter).
At equilibrium, the $\frac{1}{2}$-pseudospins align in the direction of their local fields $\bm{b}_{\bm{k}}^i=(2\Delta_{i},0,2\xi_{\bm{k}})$ in order to minimize the system's energy. 

\begin{figure}[tb]
\hspace*{-0.5cm} 
\includegraphics[width=0.5\textwidth]{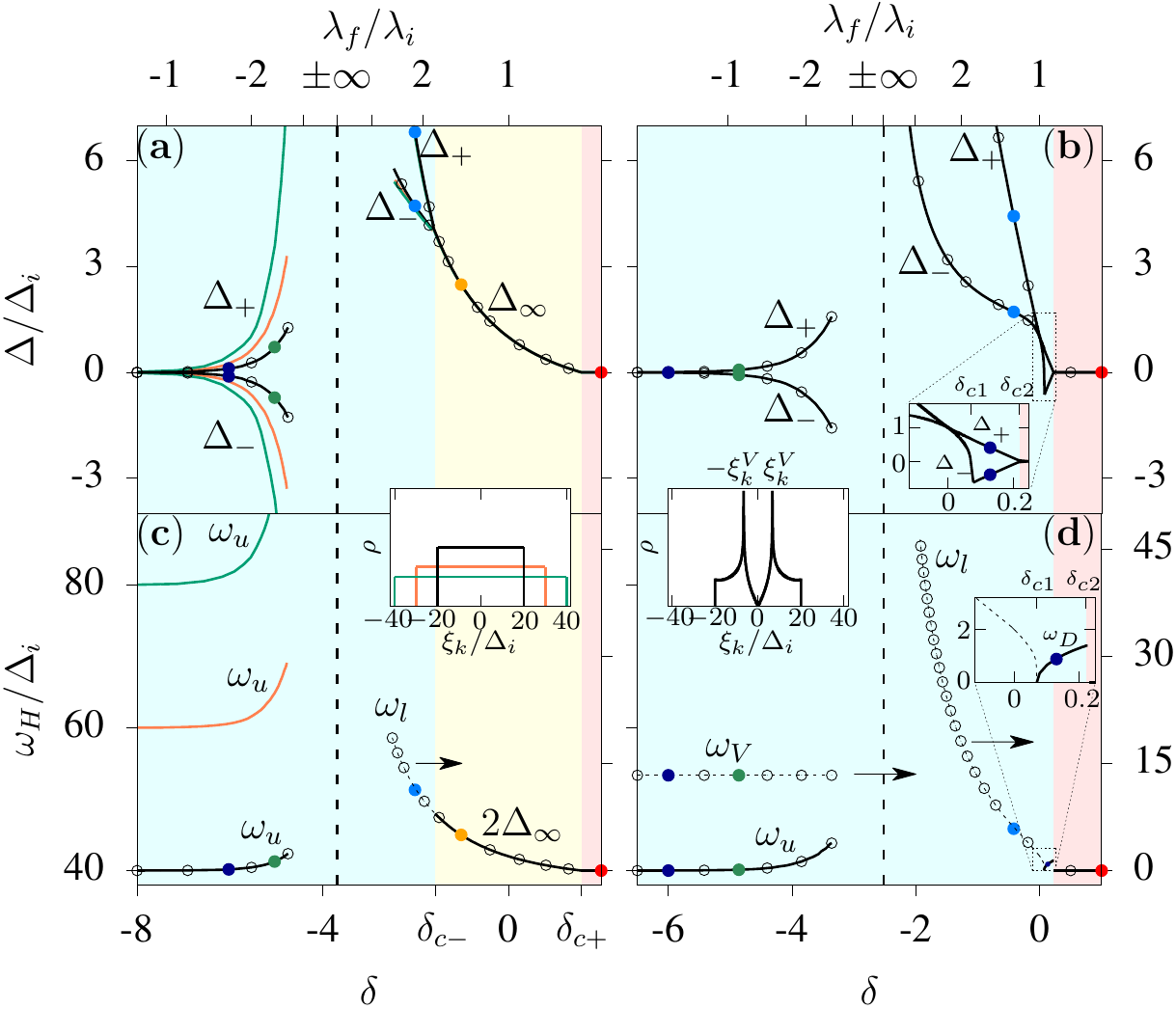}
\caption{(Color online) Dynamical phase diagrams for the constant DOS (left) and the graphene-like DOS model (right) as schematized in the white background insets. The lower scale shows the control parameter $\delta$ while the upper scale is non-linear and shows $\lambda_f/\lambda_i$ for the particular case, $W=40\Delta_i$. Panels (a) and (b) show extreme values of the order parameter characterizing the dynamics, while the lower panel shows the frequency of the Higgs oscillations labeled by the corresponding singularities in the quasiparticle DOS (see text). The background color indicates the different dynamical regimes: synchronized (magenta), damped oscillations (yellow) and overdamped (red). Full lines are obtained from the Lax root analysis, while circles are from numerical simulations (full circles correspond to dynamics shown in detail in other figures).  The dashed line fitting the $\omega_l$ frequency is $2\bar\Delta$ from the simulations, while the line is $2\Delta_\infty$ from the Lax root analysis. The dashed line fitting 
  $\omega_V$ indicates $2\xi_{\bm k}^{V}$. Both $\omega_l$  and  $\omega_V$ correspond to the right scale, as indicated by the arrows. The colored background insets are zoomed to the indicated regions. 
 } 
\label{fig:dpd}
\end{figure}

The system is prepared with an initial interaction  $\lambda_{i}$ and a gap parameter satisfying  the equilibrium gap equation,  
\begin{equation}
1=\lambda_{i}\int_{-W/2}^{+W/2}\frac{\rho\left(\xi\right)d\xi}{2\sqrt{\xi^{2}+\Delta_{i}^{2}}}
\end{equation}
where $\rho(\xi)$ is the DOS and the bandwidth satisfies, $W \gg\Delta_i$ ensuring the system is in the weak-coupling regime. The interaction is then suddenly changed to the final value, $\lambda_f$ and the gap parameter is studied as a function of time.
% This is used as initial condition and once the pairing interaction is changed,
The pseudospins evolve obeying the Bloch-like equation of motion
%%%%%%%%%%%%%%%%%%%%
%\begin{equation}
%\label{eq:eom}
$\frac{d\bm{S}_{\bm{k}}}{dt}=-\bm{b}_{\bm{k}}\left(t\right)\times\bm{S}_{\bm{k}}$
%\end{equation}
%%%%%%%%%%%%%%%%
%where we set
($\hbar\equiv1$) interacting with all the others pseudospins via the selfconsistent order parameter $\Delta(t)$.
%Refs.~\cite{Levitov2006}, we analyze the superconducting responses after a quench not only in the attractive side but also we study the phase diagram for quenches from the attractive to repulsive interactions.
We take $N=10^4, 10^5$ pseudospins within an energy range of $W=40\Delta_i, 60\Delta_i$ and $80\Delta_i$ around $\mu$. We consider both a constant DOS and the DOS of a Dirac semimetal like graphene (see white background insets of Fig.~\ref{fig:dpd}).
% We take as unit of energy the initial superconducting gap $\Delta_i$.
It is convenient to parameterize the quench by the variable  $\delta\equiv \frac{W}{\lambda_f}-\frac{W}{\lambda_{i}}$, whose value is closely related to the one already used to parametrize non-critical quenches\,\cite{Barankov2006a,Collado2019}.

The out-of-equilibrium dynamics shows collective effects and single-particle (pseudospin) excitations.
Similar to the case of periodic driving~\cite{HP2018,Collado2021,Mootz2020,Chou2017}, we find that the latter are dominated by selfconsistent pseudospin Larmor precessions encoding charge and pairing fluctuations. Their frequencies are given by $\Omega_L({\bm k})=2 E({\bm k})$ where $\pm E({\bm k})=\pm \sqrt{\xi_{\bm k}^2+{\bar \Delta}^2}$ is the quasiparticle dispersion defined in terms of an average order parameter $\bar \Delta$ computed on a large time window after the quench. The resulting DOS has edge singularities at energies $\pm\bar\Delta$ and $\pm W/2$ for quasiparticles  ($2\bar\Delta$ and $W$ for $\Omega_L$ fluctuations, with $\Delta\ll W$).

 %%%%%%%%%%%%%%%%%%%%%%%%%%%%%%%%%%%%%%%%%%%%%%%%%%%%%
%%%%%%%%%%%%%%%%%%%%%%%%%%%%%%%%%%%%%%%%%%%%%%%%%%%%%%%%

%%%%%%%%%%%%%%%%%%%%%%%%%%%%%%%%%%%%%%%%%%%%%%%%%%%%%%%%%%%%%
{\em Dynamical Phase diagram}.
%%%%%%%%%%%%%%%%%%%%%%%%%%%%%%%%%%%%%%%%%%%%%%%%%%%%%%%%%%%%%
In Fig.~\ref{fig:dpd} we present the dynamical phase diagram 
for three different bandwidths and a constant DOS (left) and for the graphene-like DOS (right). 
Panels (a) and (b) show key values of the order parameter characterizing the dynamics, while panels (c) and (d) show the generalized SHP oscillation frequencies. Full lines  were obtained exploiting the integrability of the model through a Lax roots analysis (see Supplementary Information, SI for details) and were checked by numerical simulations (circles). The frequencies of Higgs modes are labeled according to a singularity of the quasiparticle DOS to which we show below to be associated, namely, lower edge (l), upper edge (u), Van Hove (V) and Dirac point (D).

%At equilibrium the quasiparticle density of state of a BCS system has .

%%%%%%%%%%%%%%%%%%%%%%%%%%%%%%%%%%%%%%%%%%%%%%%%%%%%%
\begin{figure}[tb]
\hspace*{-0.8cm} 
\includegraphics[width=0.5\textwidth]{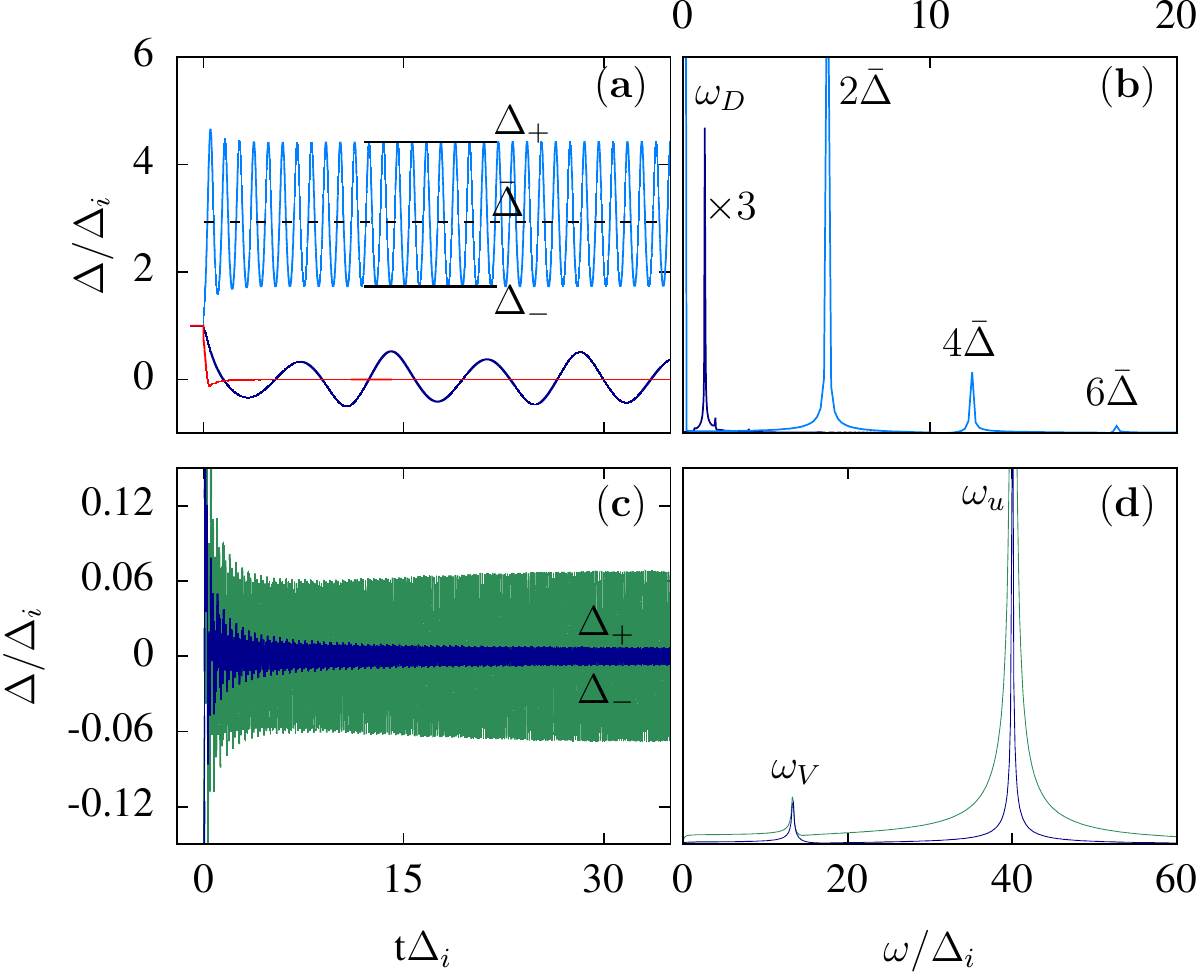}
\caption{(Color online) Representative dynamics (left column) and their FT (right column) for the graphene-like DOS case. (a) Different time dependence of superconducting order parameter for quenches in the attractive side $\lambda_f/\lambda_i>0$ (non-critical quenches). The magenta and dark-blue curves correspond to synchronized phases, while we show in red a typical dynamics in the overdamped regime. The FT are shown in the panel (b) where peaks related to Higgs mode $2\bar{\Delta}$ (and high harmonics) as well as $\omega_D$ appear. (c) Two $\Delta(t)$ for CQ $(\lambda_f/\lambda_i<0)$ and their respective FT in (d). Here, in addition to the strong response with frequency $\omega_u\simeq W$ also a frequency $\omega_{V}$ emerges. This Higgs mode is related to the Van-Hove singularities in the graphene-like DOS (see text). The quench parameters, $\delta$ used for each curve, correspond to the colored full circles of Fig.~\ref{fig:dpd}.
In all cases the FT show finite width peaks due to the finite time windows we have considered: $t\Delta_{i}\in\left[5,50\right]$ in all the cases except to resolve the slow Dirac-Higgs mode [dark blue curves in panels (a) and (b)] for which we use $t\Delta_{i}\in\left[5,300\right]$.
} 
\label{fig:dynamicsGDOS}
\end{figure}
%%%%%%%%%%%%%%%%%%%%%%%%%%%%%%%%%%%%%%%%%%%%%%%%%%%%%%%% 

In Fig.\,\ref{fig:dpd} the curves to the right of the vertical dashed line in panel (a) and (c)  reproduce the results of Ref.~\cite{Barankov2006a} where three dynamical phases were found for non-critical quenches $\lambda_f/\lambda_i>0$ (see SI Fig.~S1  %\ref{fig:dynamicsCDOS} 
for details of the dynamics at colored dots): For small increase and decrease of the attraction in the quench (small $|\delta|$), the superconducting order parameter shows damped oscillations with a Higgs frequency associated with the lower edge of the quasiparticle DOS, 
$\omega_l=2\Delta_\infty$ and saturates to a constant value $\Delta_\infty$ at long times  which therefore coincides with $\bar \Delta$ (yellow shading).
For large quenches there are two possible outcomes. 
Decreasing the pairing constant beyond a critical point ($\delta>\delta_{c+}=\pi/2$), the system goes to a gapless regime ($\Delta_\infty=0$) with an overdamped dynamics (red region). Increasing the pairing above a critical point ($\delta<\delta_{c-}=-\pi/2$), the system synchronizes and the order parameter oscillates between the values $\Delta_+$ and $\Delta_-$ with a fundamental Higgs frequency equal to twice the average order parameter\cite{Seibold2020}, $\omega_l=2\bar{\Delta}$ (magenta area). We will refer to this well known phenomena\cite{Barankov2004,Barankov2006a,Collado2019} as the ``lower-edge SHP".
%The real-time dynamics for the full colored circles can be seen in SI Fig.~\ref{fig:dynamicsCDOS}.

Using $\delta$ as the control parameter has the advantage that for large enough bandwidth, the results for $\lambda_f/\lambda_i>0$ become independent of the bandwidth, so the curves for different $W$ fall almost on top of each other. Thus, the dynamical phases obtained after non-critical quenches display a certain degree of universality, which is correctly captured in terms of the variable $\delta$. 
The upper scale (and the position of the vertical dashed line) corresponds to the particular case,  $W/\Delta_i=40$. The region close to the vertical dashed line has a final interaction in the strong coupling regime which is out of our scope, so data is omitted.

The region to the left of the vertical dashed line shows the result of the CQ. A different synchronized regime is found where the order parameter oscillates with {\em symmetric} amplitudes $\Delta_{+}$ and $\Delta_{-}$ around zero instead of having a finite average $\bar{\Delta}$ [see SI Fig.~S1(c) %\ref{fig:dynamicsCDOS}(c) 
for the detailed dynamics at the colored dots].  
This zero order-parameter average (ZOPA) behavior is reminiscent of the time-crystal phases found with periodic driving\,\cite{Collado2021}.
In contrast to the purely attractive interaction quench,
the amplitudes and the Higgs frequency are strongly dependent on the bandwidth. In this case, the SHP frequency converges to the upper edge of the fluctuation DOS ($\omega_u\rightarrow W$), when the final interaction is repulsive and small (large negative $\delta$). For $\delta<-8$ (large repulsive $\lambda_f$) the amplitude becomes vanishing small
and the dynamical phase can not be distinguished from a gapless state. Therefore, for large negative and positive $\delta$  ($\lambda_f\rightarrow \pm 0$) the system converges to the same gapless phase. 

The occurrence of SHP associated with upper and lower edges of the fluctuation/quasiparticle DOS suggests that singularities act as nucleation centers in frequency space to stabilize the synchronized phases during the non-linear dynamics. We expect this mechanism to be ubiquitous, thus leading to the appearance of Higgs mode signatures in any conventional superconducting system with singular DOS.
In order to confirm this expectation, we have studied the dynamical phase diagram of a Dirac system with a graphene-like DOS (right column) where two additional singularities are present already in the bare DOS: 
one at the Fermi level (Dirac point) and the Van Hove singularities at $\xi_{\bm k}^{V}\approx \pm 6.66\Delta_i$ (see DOS in the white background inset of Fig.~\ref{fig:dpd}). 

Interestingly, the phase diagram of the graphene-like model turns out to be quite different from the flat-DOS case. The damped regime (yellow region) is completely absent and synchronization occurs even for an \emph{arbitrary small} quench [see zoomed region inset in Fig.\,\ref{fig:dpd}(b) and the detailed dynamics at the light blue point in Fig.~\ref{fig:dynamicsGDOS}(a)]. 
Also, differently from the constant DOS model, the synchronization phenomenon takes place both for an increase and a decrease of the pairing interaction. In the latter case, decreasing enough $\lambda_f$, the pseudospins effectively decouple from each other and the gapless regime is recovered (red region). A representative dynamics of this phase is shown in Fig.~\ref{fig:dynamicsGDOS}(a) in red using the parameters indicated with red circle in Fig.~\ref{fig:dpd} (b).

Between the $\delta=0$ and the gapless phase, a quite rich transition is found, as shown in the zoomed insets of panel (b) and (d) of Fig.~\ref{fig:dpd}. First, twice the average order parameter (thin dashed line) and the
frequency $\omega_l$ decrease very rapidly tracking each other, as in other non-ZOPA synchronized phases, until a  
critical value $\delta_{c1}=0.07$ where both are driven to zero. Beyond this $\delta_{c1}$, a ZOPA synchronized phase appears associated to the Dirac point singularity with a frequency $\omega_D$ as shown in dark blue in 
Fig.~\ref{fig:dynamicsGDOS} (a) and (b). This dynamical phase is stable in a very narrow window with the frequency $\omega_D$ increasing with $\delta$ until it collapses in the overdamped phase at a second critical value $\delta_{c2}=0.22$.

The  Fourier transform (FT) of the dynamics shown in panel (b) of Fig.~\ref{fig:dynamicsGDOS} reveal that in the lower-edge SHP, the graphene-like model shows up to three harmonics while in the flat DOS model only two harmonics are visible for the present parameters [SI Fig.~S1(b)]. %~\ref{fig:dynamicsCDOS}(b)]. 
The synchronized Dirac-Higgs phase, in contrast, appears much more harmonic. 

Panel (c) and (d) of Fig.~\ref{fig:dynamicsGDOS} exemplify the dynamics for the CQ corresponding to the
matching full dots in Fig.~\ref{fig:dpd}~(b), (d), where the upper-edge SHP is excited. The overall appearance is
very similar to the case of a flat DOS [SI Fig.~S1
%\ref{fig:dynamicsCDOS}
(b), (d)]
showing ZOPA behavior. However, an extra weak modulation appears, which is revealed as a new frequency 
$\omega_V$ in the Fourier transform (d). This frequency matches twice the Van Hove singularity in the DOS $2\xi_{\bm k}^{V}$, thus as expected, a  synchronized Van Hove-Higgs mode can be excited. Its amplitude decreases in time, indicating that the mode is damped, although with quite a long coherence time,
as witnessed by the narrow peak in the FT.   
 %%%%%%%%%%%%%%%%%%%%%%%%%%%%%%%%%%%%%%%%%%%%%%%%%%%%%
\begin{figure}[tb]
\hspace*{-0.8cm} 
\includegraphics[width=0.5\textwidth]{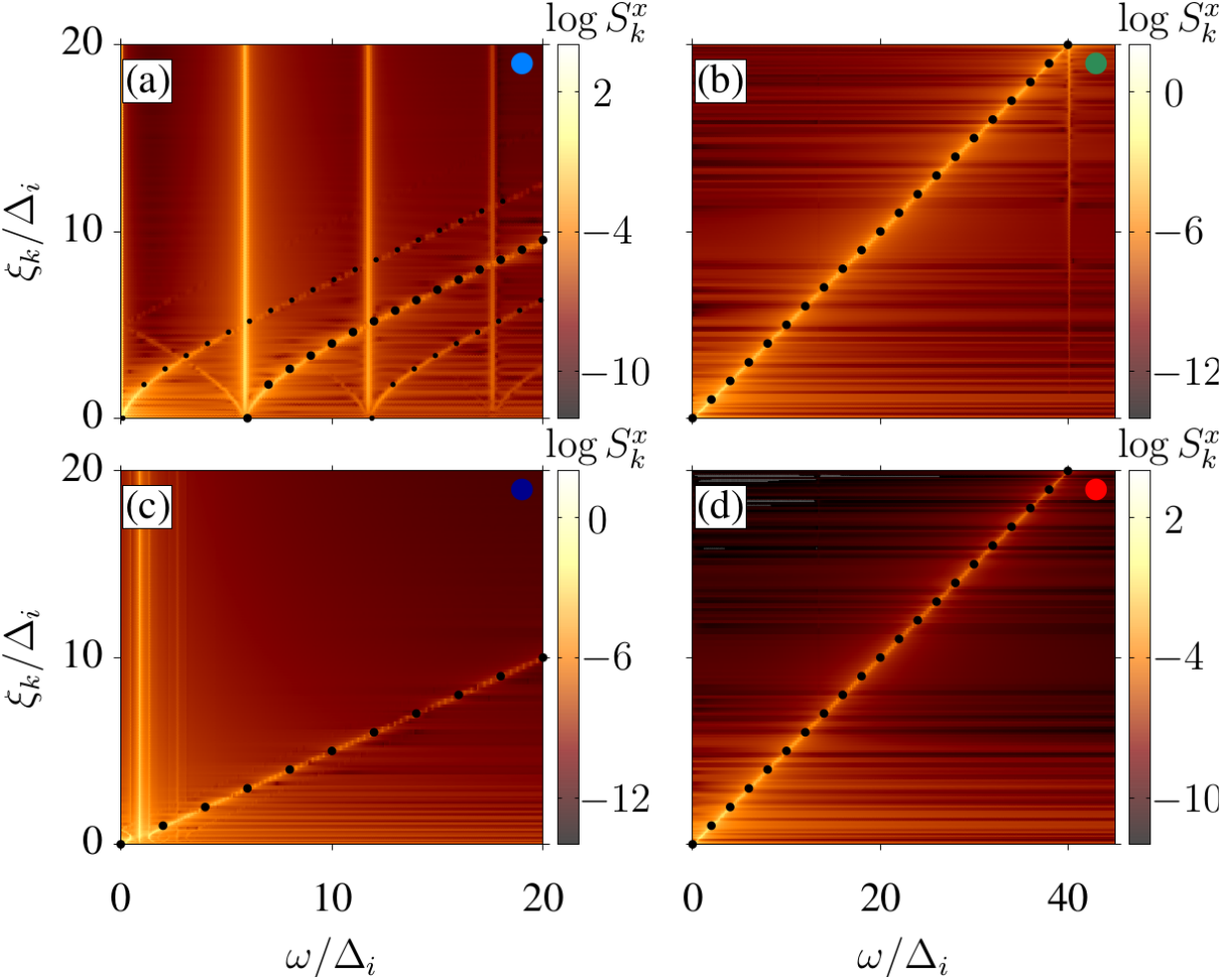}
\caption{(Color online) FT of the x-component of pseudospin texture for each of the dynamical phases assuming a graphene-like DOS. The parameter $\delta$ for each panel corresponds to the full colored circle  in Figs.~\ref{fig:dpd} and \ref{fig:dynamicsGDOS}. For all panels $\lambda_f/\lambda_i>0$ except for the right upper panel (b) in which the synchronization phenomenon (unveiled as a vertical intensity line close to $\omega=40\Delta_i$) occurs by quenching to repulsive interaction $\lambda_f/\lambda_i<0$ (CQ). The FT were computed using a time window $t\Delta_{i}\in\left[0,100\right]$.
%larger than the coherence time of Van-Hove Higgs mode. %$t\Delta_i\sim 15$.
} 
\label{fig:FFTGD}
\end{figure}
%%%%%%%%%%%%%%%%%%%%%%%%%%%%%%%%%%%%%%%%%%%%%%%%%%%%%%%% 

To fully characterizes the dynamical phases and the emergence of synchronization in the system, it is instructive to analyze the $\xi_{\bm k}$-resolved FT of the pseudospin dynamics. As we are interested in the pairing dynamics, we show the FT of the $x$-component of pseudospins  for the graphene-like model in Fig.~\ref{fig:FFTGD} (SI Fig.~S2 %\ref{fig:FFTCDOS} 
shows the same information for the flat DOS model).

%MEJOR CAMBIAR EL ORDEN. PRIMERO (c)->(a). ABAJO USO LOS LABEL ACTUALES.

Single-particle (pseudospin) excitations appear as dispersive features, while synchronized collective modes appear as vertical lines. 
For the non-ZOPA synchronized phase (light blue dots in Fig.~\ref{fig:dpd}) single-particle excitations appear in the dynamics as Larmor precessions with frequency $\Omega_L(\bm k)=2\sqrt{\xi_{\bm k}^2+\bar \Delta^2}$ as shown in panel Fig.~\ref{fig:FFTGD}(a) with the large black dots.
The same panel shows that the lower edge Higgs mode is not determined only by the quasiparticles participating in the edge singularity of $\Omega_L(\bm k)$ at the frequency $2\bar\Delta$. Indeed, the vertical feature emerging from $2\bar\Delta$ witnesses that all quasiparticles in this $\xi_{\bm k}$ window are synchronized and participate in the collective mode. Thus, the edge singularity of the dispersion can be seen as a nucleation center in frequency space given a ``rhythm'' which is followed by the rest of the quasiparticles due to the interactions.

In addition, of the main $\Omega_L(\bm k)$ dispersion, Floquet side bands appear analogous to the bands observed under periodic drive~\cite{Collado2021}. Here, of course, an external periodic drive is not present and the bands are self-generated by the action of the lower-edge Higgs mode  with frequency $\omega_l$ yielding replicas at $\Omega_L(\bm k)+ n\omega_l$ with $n=-1,0,1,2,...$ and weaker features at   $-\Omega_L(\bm k)+ n\omega_l$ with $n=0,1$.

Panel (b) shows that the upper edge singularity of the DOS is enough to trigger the SHP (appearing as a vertical feature) with frequency $\omega_u\simeq W$ provided that  the appropriate protocol, {\it i.e.} the CQ, is used. Thus, the $1/\sqrt{\omega-2\bar\Delta}$ divergence present at the lower edge of the DOS is not a prerequisite to stabilize Higgs modes.
No feature associated with the Van Hove-Higgs mode 
$\omega_V$ can be seen here, since the numerical analysis has been performed on a time window larger than its coherence time in order to have high-frequency resolution.

In panel (b) the ZOPA manifests as a gapless linear dispersion of single particle excitations,  $\Omega_L(\bm k)=2\xi_{\bm k}$ (black dots). The same linear dispersion appears for the other ZOPA modes: the Dirac-Higgs shown in panel (c) and the gapless mode shown in (d). For the former, the collective nature of the synchronization is also evident from the vertical feature at $\omega_D$.

{\em Conclusions.}
We have shown that different synchronized Higgs phases can be excited in a BCS system by choosing an appropriate quench protocol. For a given system, the frequency of the mode is determined by singularities in the DOS with small corrections due to quasiparticle interactions. The previously known lower-edge Higgs mode appears at the same frequency of a singularity in the equilibrium particle-particle response. 
The upper edge SHP is reminiscent of antibound states appearing in the equilibrium pairing response of repulsive systems. However, at equilibrium the mode is not present in particle-hole symmetric situations\,\cite{Seibold2008} while here the mode is stabilized in an out-of-equilibrium setting. Thus,  generalized Higgs modes do not appear to have always an equilibrium counterpart.

Our findings provide an innovative interpretation to the Higgs mode dynamics, which appear as synchronized quasiparticles oscillations nucleated by DOS singularities. This, in turns, implies that any spectral singularity can give rise to Higgs-mode like oscillations given a suitable quench.

The observation of the previously reported lower-edge SHP in real superconductors is hindered by its decay in other excitations and its weak coupling to light~\cite{podolsky2011visibility,scott2012rapid,Cea2015,Collado2019}. Furthermore,  signatures of Higgs dynamics in pump and probe experiments\,\cite{Matsunaga2013,Matsunaga2014,buzzi2021higgs} cannot be clearly distinguished from other Raman active modes\cite{Mansart2013} with similar frequencies, but different underlying mechanisms. The experimental detection of other Higgs modes presented here will probably encounter similar difficulties in solid state systems, as our picture is based on the BCS model, whose integrable nature does not account for thermalization.

A proper understanding of the generalized Higgs dynamics and its decay at strong coupling, as well as the possible relation with the equilibrium characteristic of the SSB phase, may be obtained by direct observation in Fermi superfluids of cold atoms. Indeed, recent improvements in the control and observation of ultracold atoms in optical lattices\,\cite{goldman2016topological,gross2017quantum} allowed the study of both equilibrium and transport properties of Fermi systems on the lattice\,\cite{mazurenko2017cold,koespell2019imaging,brown2019bad,nichols2019spin}, paving the way to the realization of the generalized Higgs dynamics described here. In particular, an artificial graphene-like lattice with tunable  interactions has been realized\,\cite{Uehlinger2013}. 
Another route is to use cold atoms in an optical cavity, which has recently been 
proposed as a BCS simulator~\cite{Lewis-Swan2021}.

Interestingly, the search for more than one Higgs boson is a subject of intense search also in high-energy scattering experiments\,\cite{Tumasyan2021}. This kind of probes, however, are more akin to equilibrium responses in condensed matter. Instead, the strongly out-of-equilibrium physics investigated here may find parallels in the electroweak transition of relevance for early universe cosmology and baryogenesis\,\cite{Ghosh2016}.

%%%%%%%%%%%%%%%%%%%%%%%%%%%%%%%%%%%%%%%%%%%%%%
\begin{acknowledgments}
We thank C. Balseiro, G. Usaj, G. Seibold, L. Benfatto, C. Castellani and M. Papinutto for useful discussions.  We acknowledge financial support from Italian Ministry for University and Research through PRIN Projects No. 2017Z8TS5B and 20207ZXT4Z and from Regione Lazio (L.R. 13/08) under project SIMAP. 
HPOC is supported by the Marie Sk\l{}odowska-Curie individual fellowship Grant agreement SUPERDYN No. 893743. This work is supported by the Deutsche Forschungsgemeinschaft (DFG, German Research Foundation) under Germany’s Excellence Strategy EXC2181/1-390900948 (the Heidelberg STRUCTURES Excellence Cluster).
\end{acknowledgments}

%\bibliographystyle{prsty_no_etal}
%\bibliography{library,library_HP}

\end{document}

\title{Supplementary Information to 
``Engineering Higgs dynamics by spectral singularities''
%  Higgs engineering in Bardeen-Cooper-Schrieffer systems
%Density-of-states engineering of Higgs mode dynamics in BCS systems
}

\author{H. P. {Ojeda Collado}}
\email{hector.pablo.ojedacollado@roma1.infn.it}
\affiliation{ISC-CNR and Department of Physics, Sapienza University of Rome, Piazzale Aldo Moro 2, I-00185, Rome, Italy}
\author{Nicol\`o Defenu}
\affiliation{Institut f\"{u}r Theoretische Physik, Eidgen\"{o}ssische Technische Hochschule Z\"{u}rich, 8093 Z\"{u}rich, Switzerland}
\author{Jos\'{e} Lorenzana}
\email{jose.lorenzana@cnr.it}
\affiliation{ISC-CNR and Department of Physics, Sapienza University of Rome, Piazzale Aldo Moro 2, I-00185, Rome, Italy}

\date{\today}
\maketitle

%%%%%%%%%%%%%%%%%%%%%%%%%%%%%%%%%%%%%%%%%%%%%%%%%%%%%%%%%%%%%

%%%%%%%%%%%%%%%%%%%%%%%%%%%%%%%%%%%%%%%%%%%%%%%%%%%%%%%%%%%%%

\setcounter{figure}{0}
\makeatletter 
\renewcommand{\thefigure}{S\@arabic\c@figure}
\makeatother

\section{Lax roots analysis}

%%%%%%%%%%%%%%%%%%%%%%%%%%%%%%%%%%%%%%%%%%%%%%%%%%%%%%%%%%%%%
%{ Lax roots analysis: Origin of Higgs modes due to singularities in the DOS}
Because of the integrability of the BCS model, the different dynamical phases described in the main text can be obtained by analyzing the integrals of motion. For this purpose, it is useful to define the so-called Lax vector \cite{Barankov2006a,Yuzbashyan2006} defined as a function of an auxiliary complex parameter $y$,
%%%%%%%%%%%%%%%%%%%%%%%%%%%%%%% 
\begin{equation}
 \label{eq:l}
 \bm{L}\left(y\right)={\bm{z}}+\lambda_f\sum_{\bm{k}}\frac{\bm{S_{\bm{k}}}}{y-\xi_{\bm{k}}}\,.
 \end{equation}
%%%%%%%%%%%%%%%%%%%%%%%%%%%%%%%%
where ${\bm{z}}$ is a unit vector along the $z$ direction,  
 and $\bm{S_{\bm{k}}}$ is the pseudospin texture before the quench. The square of the Lax vector is a conserved quantity under time evolution with the BCS Hamiltonian.
% after changing the interaction from $\lambda_i$ to $\lambda_f$.
 Therefore, the complex roots of such vector (in the following Lax roots) are also conserved. Since the square of the Lax vector is non-negative, all roots are complex-conjugated pairs. Furthermore, due to the electron-hole symmetry of the problem, it is easy to show that if the initial texture is particle-hole symmetric then if $y$ is a Lax root $-y$ is also a root. We choose the initial superconducting order parameter, $\Delta$, to be real. Also, this property is preserved at all times due to particle-hole symmetry. 

As previously discussed~[\onlinecite{Yuzbashyan2006}],  Lax roots provide information on the frequency spectrum in the Fourier transform of $\Delta(t)$ after a quench. Starting from the equilibrium pseudospin texture, a dense distribution of Lax roots appears along the real axis  which, in the thermodynamic limit, define the continuous part of the spectrum. For $t\rightarrow\infty$ this contribution vanishes in the Fourier transform of $\Delta(t)$ and only isolated pairs of complex-conjugated Lax roots (bound states) contribute corresponding to persistent oscillations. It has been shown~\cite{Yuzbashyan2006} that the number of discrete frequencies $k$ in the Fourier transform of $\Delta(t)$ is equal to $m-1$. Here, $m$ is the number of isolated pairs of complex-conjugated Lax roots. Thus, $\Delta\left(t\right)$ shows persistent oscillations at long times with $k$ different frequencies if $m>1$  while $\Delta\left(t\right)$ converges to a constant value ($\Delta\left(t\right)\rightarrow\Delta_{\infty}$) if $m=1$.

To complement the numerical results, we constructed the Lax vector and computed its roots using the equilibrium pseudospin texture for $\lambda_i$ as the initial condition.  For non-critical quenches ($\lambda_f/\lambda_i>0$) the Lax analysis was done in Refs.~\cite{Barankov2006a,Yuzbashyan2006} and the possible isolated pairs of Lax roots are always purely imaginary:  i) In the synchronized regime, there are two pairs of isolated  roots ($m=2$) namely, $y=\pm iu_1$ and $y=\pm iu_2$ that give information on the amplitude of persistent oscillation as $\Delta_\pm=(u_1\pm u_2)\Delta_i$. In this case, $\Delta(t)$ shows only one fundamental frequency ($k=m-1=1$) corresponding to the lower edge Higgs mode $2\bar{\Delta}$. ii) In the damped phase there is only one isolated pair of Lax roots ($m=1$) namely,  $y=\pm i\Delta_\infty$ indicating there are not persistent oscillations ($k=m-1=0$) but damped. iii) In the overdamped regime, the latter pair of roots collapse to the origin ($\Delta_\infty=0$) so no isolated roots are present  ($m=0$).

%Here we extend the Lax analysis also for the case of quenches to repulsive interactions $\lambda_f/\lambda_i<0$ (critical quenches CQ). 
We now extend these results to the critical quench (CQ, $\lambda_f/\lambda_i<0$) considered in the main text. 
In this case, the Lax roots are complex numbers with a \emph{finite} real part, giving precise information not only on the amplitude but also the oscillation frequency of $\Delta(t)$.
Both in the case of the constant and the graphene like-DOS, 
the roots can be written as $y=\pm (u_r\pm i u_i$) with the real part providing information on the upper edge Higgs mode frequency,  $\omega_u=2 u_r$ and the imaginary part yielding the oscillation amplitude $\Delta_{\pm}=\pm 2 u_i$. 

For the graphene-like DOS and $\lambda_f/\lambda_i>0$, there is a regime where $2\bar{\Delta}\neq 0$ for which the isolated Lax roots are purely imaginary, mimicking the non-ZOPA synchronized phase of the constant DOS model. 
Still in the synchronized regime, where the Dirac-Higgs mode  emerges, the isolated Lax roots acquire a finite real part defining the mode frequency as $\omega_D=2 u_r$ and the amplitudes of oscillations $\Delta_\pm=\pm 2 u_i$.

\section{Supplementary Figures}

%%%%%%%%%%%%%%%%%%%%%%%%%%%%%%%%%%%%%%%%%%%%%%%%%%%%%
\begin{figure}[tb]
\hspace*{-0.8cm} 
\includegraphics[width=0.5\textwidth]{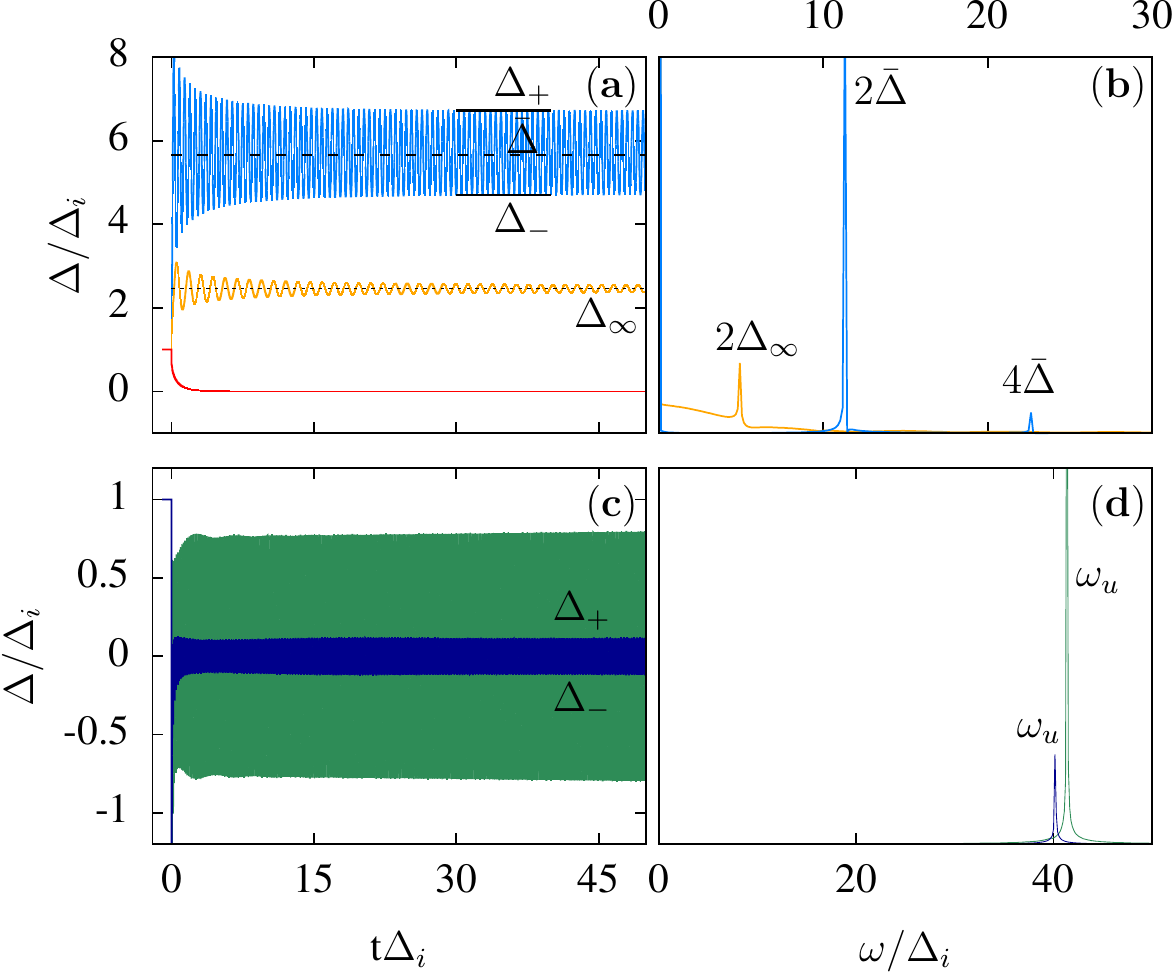}
\caption{(Color online) Representative dynamics (left column) and their FT (right column) for the constant DOS case with $W=40\Delta_i$. (a) Several time dependence of superconducting order parameter for quenches in the attractive side $\lambda_f/\lambda_f>0$. The magenta curves correspond to a synchronized dynamics while the yellow and red ones correspond to a typical damped and overdamped dynamics respectively. The FT are shown in the panel (b) where Higgs mode $2\bar{\Delta}$ and second harmonic appear. The $2\Delta_\infty$ peak corresponds to a damped mode. (c) Two $\Delta(t)$ for CQ $(\lambda_f/\lambda_i<0)$ and their respective FT in (d) showing a well-defined mode $\omega_u$ close to $W$. The quench parameters $\delta$ used in each case, have been pointed out in Fig. 1 of main text with colored dots. The FT were computed considering a time window $t\Delta_{i}\in\left[5,50\right]
$.} 
\label{fig:dynamicsCDOS}
\end{figure}
%%%%%%%%%%%%%%%%%%%%%%%%%%%%%%%%%%%%%%%%%%%%%%%%%%%%%%%% 

Figure \ref{fig:dynamicsCDOS} shows representative dynamics $\Delta(t)$ and their Fourier transform (FT) for each regime shown in the Fig.~1 (a) of the main text for the flat DOS case. The color of each curve matches the colored dots in Fig.~1 (a) of the main text encoding the simulation parameters.
Panels (a) and (b) correspond to non-critical quenches $\lambda_f/\lambda_i>0$ while (c) and (d) correspond to the CQ.
In the top panels, we can see examples of the behavior in
 the three dynamical phases previously studied in Ref.~\cite{Barankov2004} for $\lambda_f/\lambda_i>0$:
persistent oscillations (blue), damped oscillations (orange) and overdamped dynamics (red). The former corresponds to the lower-edge  Higgs mode with frequency $\omega_l=2\bar{\Delta}$ as can be seen in panel (b) where also a second harmonic peak at $4\bar{\Delta}$ appears. In the damped regime, a peak at frequency $2\Delta_\infty$ is also resolved due to the finite time window we consider computing the FT. 
For the CQ (panels (c) and (d)), similar to the graphene-like DOS case discussed in the main text, $\Delta(t)$ shows persistent oscillations between $\Delta_\pm$ with the upper-edge Higgs mode frequency $\omega_u$.

  %%%%%%%%%%%%%%%%%%%%%%%%%%%%%%%%%%%%%%%%%%%%%%%%%%%%%
\begin{figure}[t]
\vspace*{.2cm} 
\includegraphics[width=0.47\textwidth]{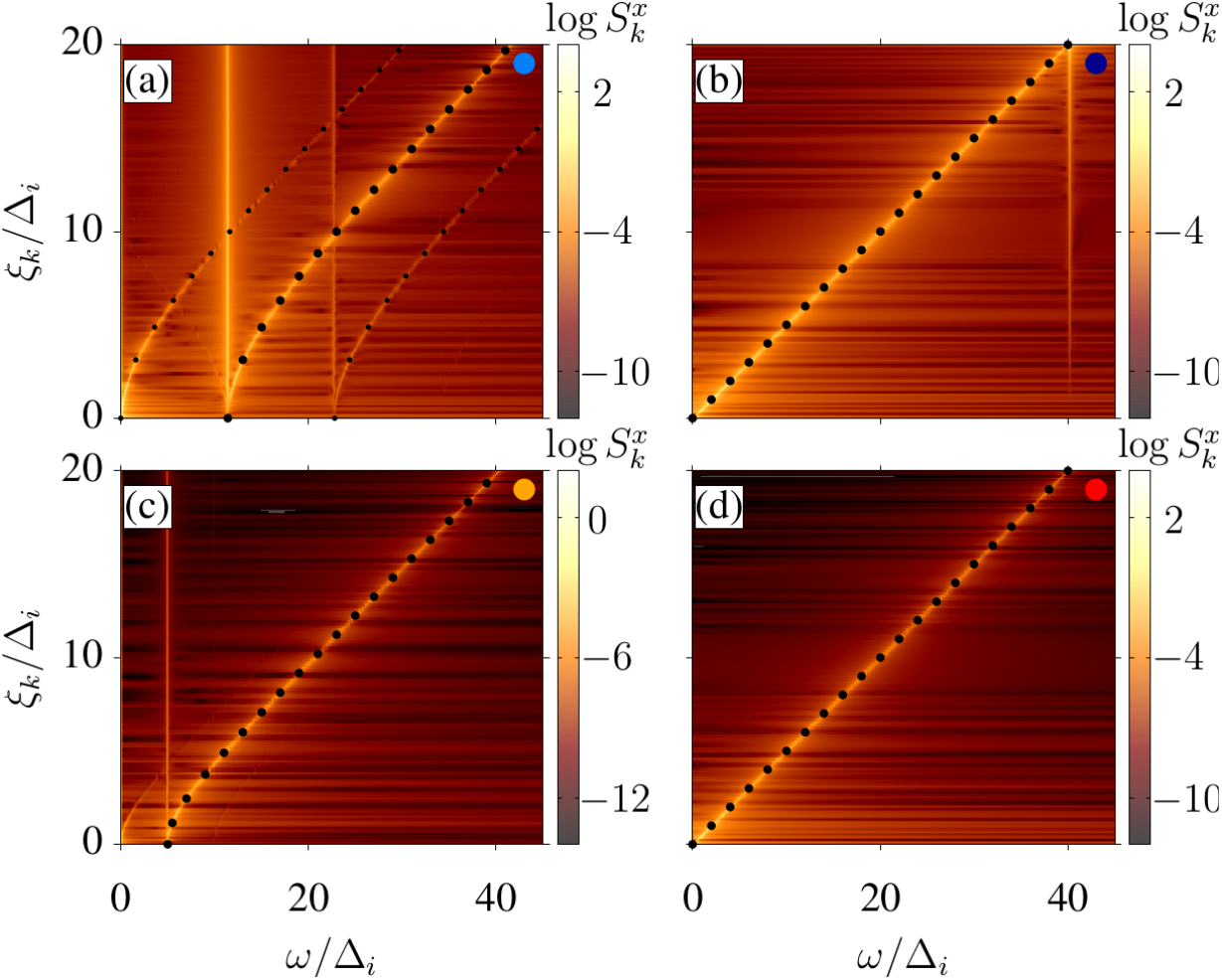}
\caption{(Color online) Fourier Transforms of the $x$-component of the pseudospin textures for the four possible dynamical phases in the constant DOS model. For each panel, the used quench parameters ($\delta$) corresponds to the matching full colored circles  in Fig.~1 of the main text and the colors of the dynamics in Fig.~\ref{fig:dynamicsCDOS}. For all panels $\lambda_f/\lambda_i>0$ except for (b) which corresponds to the CQ.  The synchronization phenomenon appears as a vertical intensity line close to the upper edge, $\omega=40\Delta_i$.  The FT were computed considering a time window $t\Delta_{i}\in\left[0,100\right].
$}
\label{fig:FFTCDOS}
\end{figure}
%%%%%%%%%%%%%%%%%%%%%%%%%%%%%%%%%%%%%%%%%%%%%%%%%%%%%%%% 

 Figure Fig.~\ref{fig:FFTCDOS} (a) shows the pseudospin resolved FT of the dynamics in the synchronized phase with $\lambda_f/\lambda_i>0$. The large black dots correspond to the Larmor  frequency $\Omega_L=2\sqrt{\xi_{\bm k}^2+\bar{\Delta}^2}$. In addition, the spectrum shows Floquet satellites at $\Omega_L\pm \omega_l$. The vertical features show that all pseudospins respond with the Higgs frequency $2\bar{\Delta}$ and high harmonics. 
 
Figure~\ref{fig:FFTCDOS} (c) shows the same information  
in the damped regime. Each pseudospin oscillates with an effective Larmor frequency $2\sqrt{\xi_{\bm k}^2+\Delta_\infty^2}$ which introduces dephasing giving rise to damped oscillations in $\Delta(t)$ [c.f. Fig.~\ref{fig:dynamicsCDOS} (a)]. The vertical line at $\omega=2\Delta_\infty$ is due to the finite time window in which the FT is performed. In the long time limit, its spectral weight vanishes, indicating the absence of persistent oscillations.

 By decreasing enough the final coupling, the system enters into the gapless phase, where each pseudospin oscillates  with its own bare Larmor frequency $2\xi_{\bm k}$  [panel (d)].

 In the case of the CQ [panel (b)] the pseudospin resolved FT shows both the individual Larmor frequency, $2\xi_{\bm k}$ and a vertical feature in the upper edge corresponding to the upper-edge Higgs mode discussed in the main text for a constant DOS. The frequency is  $\omega_u\approx W=40\Delta_i$ and the intensity increase with  $\xi_{\bm k}$ indicating that high energy pseudospin participate with larger amplitude.

%\bibliographystyle{prsty_no_etal}
%\bibliography{library,library_HP}